**Major Article**

# Coexistence under hierarchical resource exploitation: the role of *R**-preemption tradeoff

Man Qi[1,2φ*] (manqi@ ccnu.edu.cn)

Niv DeMalach[3 φ] (niv.demalach@mail.huji.ac.il)

Hailin Zhang[2](hailzhang@mail.ccnu.edu.cn)

Tao Sun[2*](suntao@bnu.edu.cn)

φ These authors contributed equally to this work

* Corresponding author: Man Qi, Tel: 86-18612595198; Tao Sun, Tel:86-10-58805053, Fax:86-10-58805053

1 Key Laboratory of Geographical Process Analysis & Simulation of Hubei Province/College of Urban and Environmental Sciences, Central China Normal University, Wuhan 430079, China

2 State Key Laboratory of Water Environment Simulation, School of Environment, Beijing Normal University, Beijing 100875, China

3 Institute of Plant Sciences and Genetics in Agriculture, Faculty of Agriculture, Food, and Environment, The Hebrew University of Jerusalem, Rehovot, Israel

## ABSTRACT

Resource competition theory predicts coexistence and exclusion patterns based on species' $R^*$s, the minimum resource values required for a species to persist. A central assumption of the theory is that all species have equal access to resources. However, many systems are characterized by preemption exploitation, where some species deplete resources before their competitors can access them (e.g., asymmetric light competition, contest competition among animals). We hypothesize that coexistence under preemption requires an $R^*$-preemption tradeoff, i.e., the species with the priority access should have a higher $R^*$ (lower 'efficiency'). Thus, we developed an extension of resource competition theory to investigate partial and total preemption (in the latter, the preemptor is unaffected by species with lower preemption rank). We found that an $R^*$-preemption tradeoff is a necessary condition for coexistence in all models. Moreover, under total preemption, the tradeoff alone is sufficient for coexistence. In contrast, under partial preemption, more conditions are needed, which restricts the parameter space of coexistence. Finally, we discussed the implications of our finding for seemingly distinct tradeoffs, which we view as special cases of $R^*$-preemption tradeoff. These tradeoffs include the digger-grazer, the competition-colonization, and tradeoffs related to light competition between trees and understories.

## INTRODUCTION

Explaining the tremendous diversity of plants and animals is a major challenge in ecology (Gauze 1934; Hutchinson 1959; Hardin 1960; Tilman 1982; Chesson 2000; Pennisi 2005; Vellend 2016). Classical resource competition theory (Volterra 1928; MacArthur 1972; Armstrong and McGehee 1980; Tilman 1982) provides a simple mean for predicting diversity patterns from a mechanistic description of resource exploitation. This theory predicts patterns of competitive exclusion and coexistence based on species' $R^*$s, the minimum resource values required for a species to persist (also equivalent to the resource levels in a monoculture at equilibrium, Tilman 1982).

A central assumption of resource competition theory is *equal-access exploitation*, i.e., all species have identical access to resources (Volterra 1928; MacArthur 1972; Armstrong and McGehee 1980; Tilman 1982). Equal-access exploitation (also known as 'scramble competition' or 'symmetric competition') is a reasonable approximation for well-mixed systems such as algae species in an aquatic ecosystem (Tilman 1977). However, some systems are characterized by preemption exploitation ('contest competition' or 'asymmetric competition') where species deplete their resources before their competitors can access them (Schoener 1976; Grime 1977; Huston and DeAngelis 1994; Schwinning and Weiner 1998; Craine and Dybzinski 2013; DeMalach *et al.* 2016). Common examples include fast-moving animal species that deplete resources before slow-moving competitors arrive (Richards et al. 2000; Kneitel and Chase 2004) and tall plants that deplete light for shorter plants (Tilman 1988; Schwining and Weiner 1998; DeMalach *et al.* 2016).

Regardless of the specific mechanisms by which species are competing for resources, as long as species get equal access to a resource, coexistence requires each species to specialize on a different resource, i.e., a tradeoff in $R^*$s (Tilman 1982). Here, we propose that coexistence under preemption requires a tradeoff between preemption rank and $R^*$. In a simple two-species scenario, it implies that one species has a lower $R^*$ (hereafter gleaner), while the other has priority in accessing the resource (hereafter preemptor). We argue that seemingly distinct tradeoffs among organisms can be viewed as specific cases of this tradeoff.

The grazer-digger tradeoff (also known as 'exploiter-explorer' or 'cream-skimmer—crumb-picker') suggests that animals with fast movement (i.e., grazers) can reach the resource preemptively before slow-moving species (i.e., diggers) arrive. However, the more efficient diggers can persist on the leftovers (Richards *et al.* 2000; Kneitel and Chase 2004). According to the body-size tradeoff (Basset 1995), large animals are more aggressive, which allows them to preempt resources, but smaller species are more efficient foragers. Likewise, the dominance-discovery tradeoff in ants (Adler *et al.* 2007) suggests that foraging in large groups allows preemption (by aggressive behavior) while foraging efficiency is higher when ants have more scattered spatial spread. Similarly, the $R^*$- preemption tradeoff lies at the heart of many tradeoffs among plants, such as the height tradeoff (Givnish 1982; Falster and Westoby 2003; Onoda *et al.* 2014) and leaf-economy tradeoff (Onoda *et al.* 2017).

Here, we investigated the effect of $R^*$- preemption tradeoff on coexistence and relative abundance patterns under the different types of preemption. First, we examine a model with *total preemption*, where the gleaner uses only the leftover of the better preemptor and therefore does not affect the latter. This model describes competition for light between trees and understory herbs. Then, we investigate a model with *partial preemption,* where some portion of the resource is available only to the preemptor while the rest is available to both species. The second model resembles a grazer-digger tradeoff among animals where a preemptor can partially preempt the resource due to its faster discovery rate. Still, once discovered by the gleaner, the resource is accessible to both species.

## METHODS

Our models follow the simplifying assumptions of the classical models (Tilman 1982), including no age structure and lack of spatial and temporal heterogeneity in environmental conditions. Although simple, our models are mechanistic in the sense that all interactions are mediated by resource depletion instead of assuming 'direct interactions' as in phenomenological models (see Letten *et al.* 2017, for details).

For simplicity, we examined a scenario of a single type of resource (e.g., light, specific food type) and two competitors. However, we show that under total preemption, an arbitrary number of species can coexist under the $R$*-preemption tradeoff (AppendixS1 and S2).

### A light competition model

In this model, the preemptor is a canopy tree (species 1), and the gleaner is an understory herb (species 2). The following function describes the population dynamics of both species:

$$(1)\quad \frac{dN_i}{dt} = (a_i w_i I_i - m_i)N_i$$

Where $N_i$ is population density, $m_i$ is the mortality rate, $a_i$ is the resource depletion rate, and $w_i$ is a conversion factor of light depletion to population growth. $I_i$ is the amount of unused light experienced by the $i^{th}$ species, i.e., the unused light determines population growth. Light levels are denoted by $I$ rather than $R$ to highlight that we describe a flux rather than concentration as for most other resources (following Dybzinski and Tilman 2007).

The amount of unused light experienced by species $1(I_1)$ is directly related to its density:

$$(2)\quad I_{1(t)} = \max(g - a_1 N_{1(t),} 0)$$

where $g$ is the resource influx, and $a_1N_1$ is light consumption by established trees. The amount of unused light available for the population growth of understory herbs ($I_2$) equals the total light influx (g) minus light consumption by established adults of its own species ($a_2N_2$) and that of the trees ($a_1N_1$):

(3) $I_{2(t)} = \max(g - a_1N_{1(t)} - a_2N_{2(t)}, 0)$

**A grazer-digger model**

In this model, the preemptor is an animal with a fast movement (grazer), and the gleaner is an animal with more efficient resource use (digger). The model includes a 'resource' ($R_1$) which is available only to the grazer (have not been discovered by the digger), and a 'resource' ($R_2$) which is available to both species. Population growth in this model is affected by resource concentration:

(4) $\frac{dN_1}{dt} = [a_1w_1(R_1 + R_2) - m_1]N_1,$

(5) $\frac{dN_2}{dt} = (a_2w_2R_2 - m_2)N_2,$

The following equations describe the dynamics of the two resources:

(6) $\frac{dR_1}{dt} = g - a_1R_1N_1 - dR_1 - qR_1$

(7) $\frac{dR_2}{dt} = dR_1 - a_1R_2N_1 - a_2R_2N_2 - qR_2$

where $g$ is the resource influx of $R_1$, $a_1R_1N_1$ is resource consumption by the preemptor, $d$ is the rate by which $R_1$ is 'transforming' into $R_2$, i.e., the discovery rate of the gleaner). Unlike $R_1$, $R_2$ is consumed by both species ($a_1 R_2 N_1$, $a_2 R_2 N_2$). The decay rate ($q$) is assumed to be equal for both resources

**Table 1. Variables and parameters of the models.** The subscript *i* indicates species-specific (except for $R_i$) variable\parameter values.

| Symbol | Description | Type |
|---|---|---|
| $N_i$ | Population density | Variable |
| $R_i$ | Resource concentration (grazer-digger model) | Variable |
| $I_i$ | Unused light level (light competition model) | Variable |
| $m_i$ | Mortality rate | Parameter |
| $a_i$ | Resource depletion rate | Parameter |
| $w_i$ | Conversion factor from resource depletion to population growth | Parameter |
| $g$ | Resource influx | Parameter |
| $q$ | Resource decay rate (grazer-digger model ) | Parameter |
| $d$ | Discovery rate of the gleaner $R_i$ (grazer-digger model ) | Parameter |

## RESULTS

### Light competition model

To be consistent with the classical terminology of $R^*$ in Tilman (1982), we define $I^*$ of each species as the amount of unused light leading to zero net growth (or the amount of unused light of a monoculture). Thus, $I^*$ can be calculated by assuming a quasi-steady approximation of $\frac{dN_i}{dt} = 0 \rightarrow (a_i w_i I_i - m_i) N_i = 0$:

$$(8) \quad I_i^* = \frac{m_i}{a_i w_i}$$

We define the unused light in the absence of consumers as $I_1^0 = I_2^0 = g$. An abiotic extinction (i.e., when the population growth rate is negative in the absence of competitors) occurs if $I_i^* > I_i^0$. When both species' values are lower than $I_i^0$, coexistence and exclusion are determined by the difference among species in their $I^*$ values (Fig 1, Appendix S1). When $I_1^* < I_2^*$, the tree excludes the herb. When $I_1^* > I_2^*$, coexistence is possible because the herb can persist on the leftover of the tree. This rule can be generalized to any number of species: a species with a lower preemption rank and lower $I^*$ than resident species can always invade a community in equilibrium, and since the invader cannot affect the resident species, a theoretically infinite number of species can coexist on a single type of resource under total preemption (Appendix S1).

Patterns of relative abundance are more complex than coexistence (Fig 1, Appendix S1). They are affected by both species' traits ($I^*$) and resource availability ($I^0$). The condition for herbs to exceed the abundance of the trees is:

$$(9) \quad \frac{m_2}{a_2 w_2} < \left(1 + \frac{a_2}{a_1}\right) \frac{m_1}{a_1 w_1} - g \frac{a_2}{a_1} \quad \text{and} \quad \frac{m_1}{a_1 w_1} < g$$, which can be rewritten as:

$$(10)\quad I_2^* < \left(1+\frac{a_2}{a_1}\right) I_1^* - g\,\frac{a_2}{a_1} \quad \text{and} \quad I_1^* < g$$

These equations imply that an increase in resource influx ( $g$ ) increases the relative abundance of the preemptor but does not affect the gleaner (see Appendix S1: equations S1~S4).

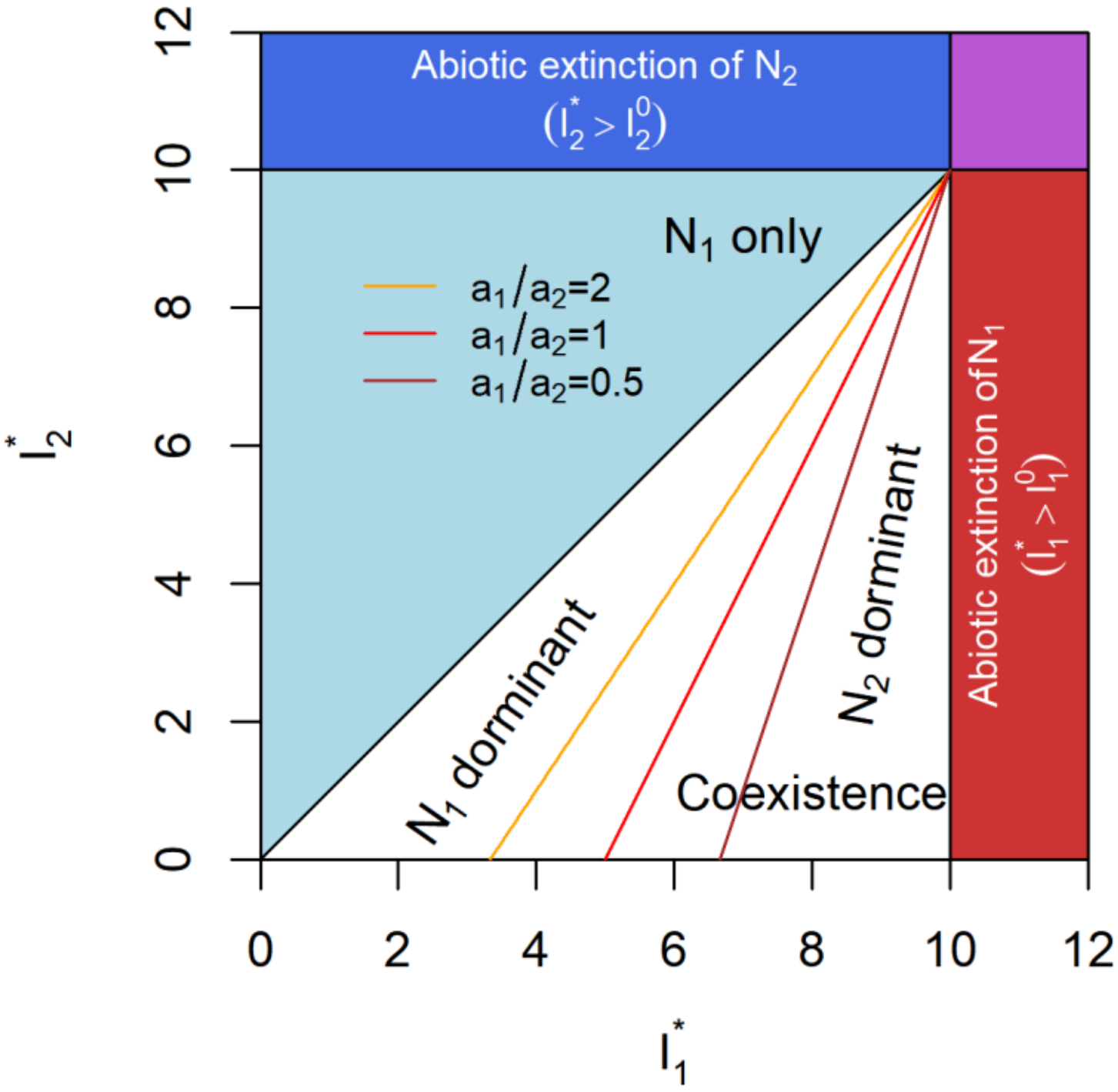

**Fig. 1. Competitive outcomes for tree\preemptor (*$N_1$*) and an understory herb species\gleaner (*$N_2$*) in light competition model (total preemption). The four qualitative outcomes: abiotic extinction (for any species with $I^*$ higher than $I^0$), competitive exclusion ( $I_1^* < I_2^* < I^0$ ), and coexistence ( $I_2^* < I_1^* < I^0$ ). $I^0 = g = 10$. The coexistence region is divided based on the relative abundance of the species. The lines represent the transition between a higher abundance of trees (the left side of each line) and a higher abundance of understory herbs (the right side). Different lines represent different parameter values based on equation 10.**

## Grazer-digger model

We define here $R^*$ for each species as the resource level of zero net growth (equivalent to the equilibrium resource concentration of a monoculture, sensu Tilman 1982). Thus, while the biological interpretation of $I^*$ and $R^*$ is different (Dybzinski & Tilman 2007), they are equally affected by the same parameters $\frac{dN_i}{dt} = 0 \rightarrow (a_i w_i R_i - m_i) N_i = 0$:

$$(11) \quad R_i^* = \frac{m_i}{a_i w_i}$$

Furthermore, the conditions for abiotic extinctions are similar though quantitatively different. An abiotic extinction of $N_1$ occurs when $R_1^* > R_1^0$ , where $R_1^0$ is the steady-state resource concentration available for grazers in the absence of consumers, and $R_1^0 = g / q$ (i.e., $R_1^{(0)} + R_2^{(0)}$ see Appendix S2: equation S2-1). An abiotic extinction of $N_2$ occurs when $R_2^* > R_2^0$ , where $R_2^0$ is the steady-state resource concentration available for diggers in the absence of consumers, and $R_2^0 = \frac{dg}{(q+d)q}$ ( Appendix S2: equation S2-1).

Nonetheless, the conditions for exclusion and coexistence are qualitatively different for partial preemption exploitation. The gleaner can exclude preemptor under the condition $R_2^* \le R_1^* - \frac{g}{d+q}$, limiting the parameter space of coexistence to

$$R_1^* - \frac{g}{d+q} < R_2^* < \frac{d\left(R_1^*\right)^2}{(dR_1^* + g)} \quad \text{(Fig. 2).}$$

The condition for gleaner to exceed the abundance of the preemptor is (Appendix S2: equation S2-5):

(12) $$R_2^* < \frac{-\left[\left(\frac{a_1}{a_2}+1\right)g+\left(\frac{a_1}{a_2}d-d-q\right)R_1^*\right]+\sqrt{\left[\left(\frac{a_1}{a_2}+1\right)g+\left(\frac{a_1}{a_2}d-d-q\right)R_1^*\right]^2+4d\left(d+q\right)\frac{a_1}{a_2}\left(R_1^*\right)^2}}{2(d+q)}$$

These equations imply that an increase in resource influx ($g$) increases the relative abundance of preemptor, and an increase in its discovery rate ($d$) increases the relative abundance of the gleaner. Therefore, while an increase of $q$ decreases the abundance of preemptor, it does not affect the gleaner when preemptor exists. (see details in Appendix S2, equation S3-4).

In sum, the light competition and grazer-digger models differ in two interrelated aspects. First, only in the grazer-digger model, the gleaner can exclude the preemptor. Second, the coexistence region is much smaller in the grazer-digger model, where the R*-preemption tradeoff is insufficient for coexistence.

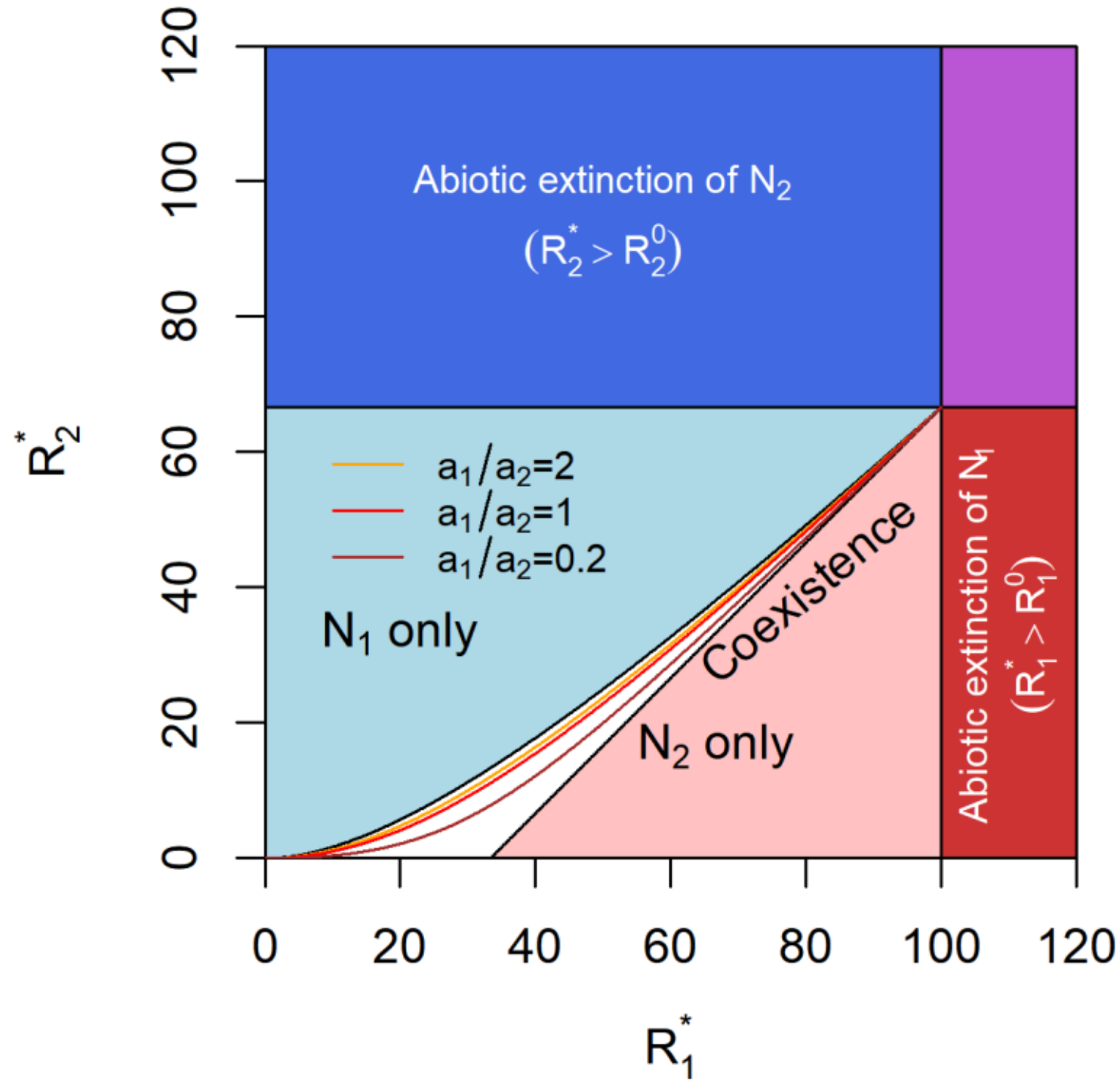

**Fig. 2. Competition outcome for grazer\preemptor (*$N_1$*) and digger/gleaner (*$N_2$*). The four qualitative outcomes are abiotic extinction (for any species with $R^*$ higher than $R^0$), competitive exclusion of N1 ($R_2^* \le R_1^* - \frac{g}{d+q}$ and $R_1^* < R_1^0$), competition exclusion of N2 ($R_2^* \ge \frac{d\left(R_1^*\right)^2}{(dR_1^* + g)}$ and $R_2^* < \frac{dg}{(q+d)q}$), and coexistence ($R_2^* < \frac{d\left(R_1^*\right)^2}{(dR_1^* + g)}$ and $R_2^* > R_1^* - \frac{g}{d+q}$). The coexistence region is divided based on the relative abundance of the species. The lines represent the transition between a higher abundance of the grazer (the upper side of each line) and a higher abundance of digger (the lower side). Different lines represent different $a_1/a_2$ values based on equation 12. g=10, d=0.2, q=0.1 in the graph.**

## DISCUSSION

Within our models, the tradeoff between preemption ability and $R^*$ is a necessary condition for coexistence under preemption exploitation. Furthermore, in the light competition model, this tradeoff alone is sufficient for coexistence. In contrast, coexistence in the grazer-digger model requires additional conditions because the gleaner species can exclude the preemptor at some parameter spaces.

We attribute the differences between the two models to differences between partial and total preemption because the other difference between the models (concentration vs. flux resource types) has a negligible effect on the outcomes (Appendix S3). Below, we discuss the implications of our model for different organisms and the new questions arising from our findings.

### Preemption in plant communities

Asymmetric competition for light is a classic example of preemption exploitation in plants (Schwining and Weiner 1998). Theoretical (DeMalach *et al.* 2016) and empirical studies (Lamb *et al.* 2009; DeMalach *et al.* 2017; Hautier *et al.* 2018) demonstrate that transition from relatively symmetric competition belowground to asymmetric competition for light is a major driver of species loss in various grasslands. Many grassland species are not efficient enough to grow under low light levels, and therefore preemption competition results in the extinction of short-statured species (Dickson *et al.* 2014; DeMalach *et al.* 2017). More broadly, in the absence of $R^*$-preemption tradeoff, preemption exploitation can lead to lower diversity than equal-access exploitation because of large ecological fitness differences (Sensu Chesson 2000).

However, in many well-established forests, there is a layer of understory herbs that persist despite light preemption by trees (De Frenne *et al.* 2013). We use these phenomena as an example to illustrate how the model's parameters relate to biological attributes (Fig. 3). The primary condition for coexistence between trees and understory species is that the understory species (gleaner) will have a lower $R^*$ than the trees. Thus, coexistence is impossible in a dense forest without highly efficient understory species (Fig. 3a). In contrast, in some systems, high light capture efficiency (high $a$), and ability to perform photosynthesis under low light conditions (high $w$) of understory species (Valladares *et al.* 2002) enables them to survive on the unused light (Fig. 3b). Additionally, a low population density of trees (as affected by $m$, $w$) may allow sufficient radiation for the persistence of herbaceous vegetation (Fig. 3c). Coexistence can also occur when the per-capita effect of trees on light availability ($a$, which is closely related to leaf area index) is small (Fig. 3d).

The $R^*$-preemption tradeoff can be relevant to the coexistence of species varying in phenology. For example, in the Mediterranean annual communities of California, some species grow and deplete water much earlier than others (Godoy and Levine 2014; Alexander and Levine 2019). Theoretically, a late-phenology species can persist on the leftovers of the early phenology species by withstanding lower water potential (lower $R^*$). However, since there is often an overlap of the growing seasons, it implies partial (rather than total) preemption, which restricts the conditions for coexistence. Furthermore, empirical evidence suggests that phenology also affects many other processes. Therefore, it is difficult to merely attribute coexistence patterns among species varying in phenology to the parameters of our model (Godoy and Levine 2014; Alexander and Levine 2019).

One of the most common explanations for coexistence among plants and other sessile organisms is a tradeoff between competition and colonization abilities (Hastings 1980; Tilman 1994; Calcagno *et al.* 2006). Interestingly, this tradeoff could also be viewed as a special case of the $R^*$-preemption tradeoff where space is the limiting resource, and $R^*$ is the fraction of empty patches in a monoculture at equilibrium. The 'better competitor' is analogous to the preemptor in the model of total preemption because it is not affected by the other species. Additionally, the colonizer without priority access to space is restricted to preemptor's leftovers (empty patches). In these models, coexistence requires the colonizer to have a lower '$R$*' than the preemptor in the sense that its monoculture should have fewer empty patches at equilibrium.

Although we propose that the competition-colonization tradeoff is a particular case of the preemption-$R^*$ tradeoff, the specific models of competition-colonization tradeoff (Hastings 1980; Tilman 1994; Calcagno *et al.* 2006) are different from our models because space is not consumed similarly to other resources. Moreover, within our model, $R$* and $I$* are related to resource depletion ($a$) and conversion ($w$) rather than dispersal or fecundity as in the competition-colonization models.

**Preemption in animal communities**

Several tradeoffs that enable coexistence on a single *resource typ*e were proposed for animal communities, including the body-size tradeoff, grazer-digger tradeoff, and dominance-discovery tradeoff (Basset 1995; Richards *et al.* 2000; Adler *et al.* 2007; Basset and DeAngelis 2007). We argue that all these tradeoffs are special cases of the preemption-$R^*$ tradeoff.

More broadly, animal ecologists have long recognized that preemption exploitation ('contest competition') is a common phenomenon in animal communities (Nicholson 1954; Lawton and Hassell 1981). However, so far, contest competition was not incorporated into 'mechanistic models' and was only described by 'phenomenological models' where the effects of competition are assumed to be direct instead of mediated by resource consumption (Hassell 1975; Brannstrom and Sumpter 2005). Mechanistic models provide a better understanding of the underlying processes Letten *et al.* 2017; Koffel *et al*. 2021). Additionally, the number of parameters scales linearly rather than quadratically with species number in mechanistic models, making them less data demanding for multispecies communities (Tilman 1982).

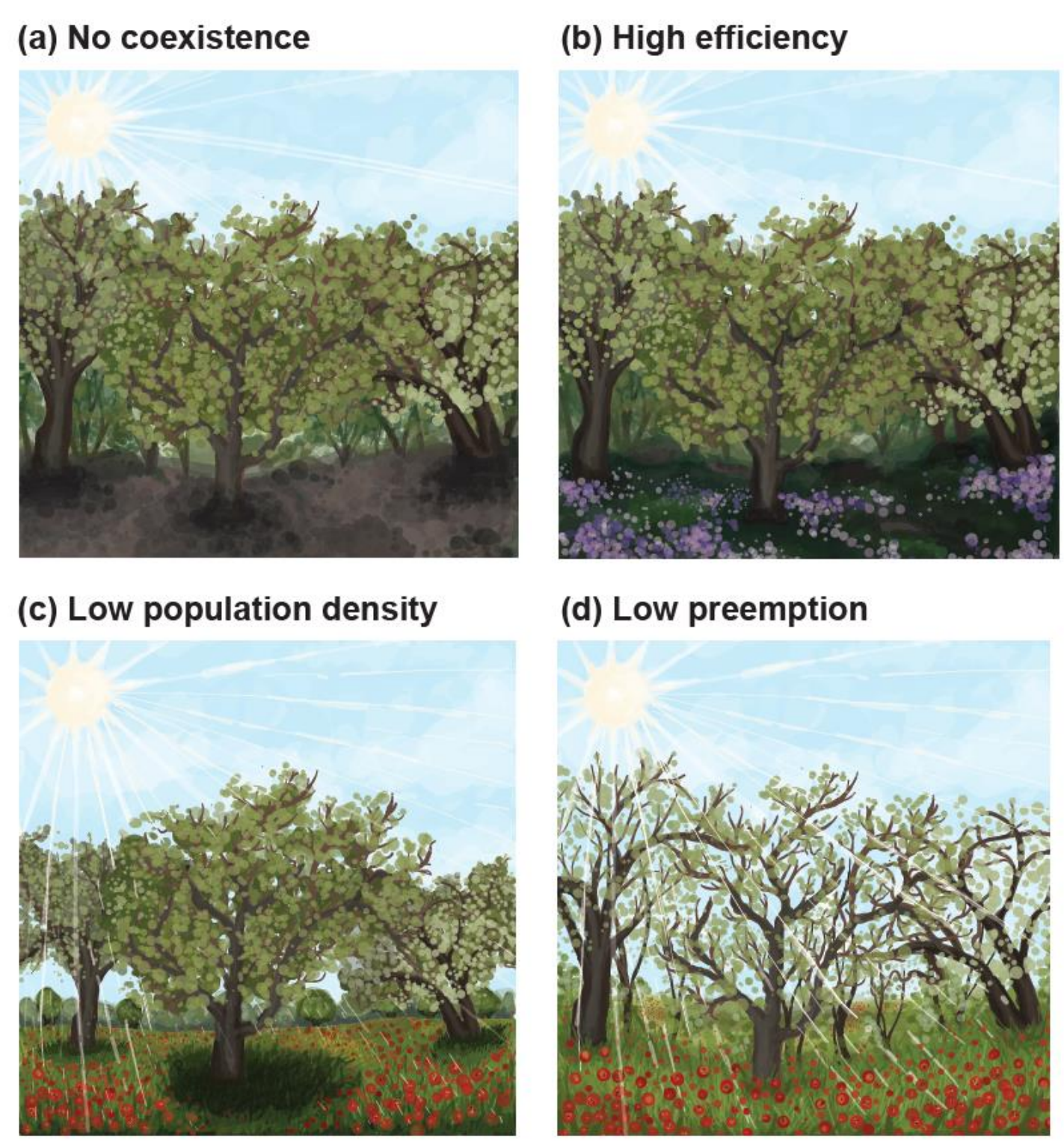

**Fig. 3 An illustration of the conditions for coexistence for trees (preemptor) and understory herbs (gleaner) competing for light. (a) A dense canopy of trees (as affected by *m*, and *w*, and *a*) can prevent the growth of understory species by reducing light availability. (b) Understory species can persist despite low light availability when having high conversion efficiency (*w*). (c) Low tree density (higher *m* or lower *w* of tree) can enable the persistence of understory species in the gaps. (d) Low per-capita depletion (*a*) by the trees (driven by low canopy density here) increases the *trees' I**, thereby enabling the persistence of understory herb.**

## Comparison between preemption and equal-access exploitation

Under equal-access exploitation, a single *type* of limiting resource does not permit coexistence ('the competitive exclusion principle', Voltera 1928; Hardin 1960; McArthur and Levins 1964; Levin 1970; Armstrong and McGehee 1980). However, under preemption exploitation, each species experiences different resource availability, and therefore, the number of *effective* resources always equals the number of species.

Notably, our findings do not refute the competitive exclusion principle but demonstrate that what is a single resource under equal-access exploitation (e.g., light) effectively acts as distinct resources under preemption exploitation, thereby allowing more opportunities for coexistence (see Abrams 1988, for a thorough discussion on how resources should be counted).

The classical theory assumes that all species experience equivalent levels of resource. In contrast, our model assumes a hierarchy of resource acquisition, where species with lower preemption ranks have better access to resources. This hierarchy can be strict where the preemptor is not affected by other species (total preemption) or not (partial preemption). Partial preemption lies between total preemption and completely equal-access exploitation.

The continuum between preemption and equal-access exploitation was investigated in animal ecology using phenomenological models (Hassell 1975). Still, equal-access exploitation was assumed when seeking a mechanistic explanation for the models (Geritz and Kisdi 2004). In plant ecology, the continuum was described within the contest of size-asymmetry, a quantitative measure of the degree of size-related differences in resource acquisition (Schwining and Weiner 1998; DeMalach *et al.* 2016). Notably, the continuum between equal-access exploitation and hierarchical preemption is not necessarily affected only by the size nor restricted to plants. We, therefore, hope that future mechanistic models will investigate this continuum in various other contexts.

## Concluding remarks

In this contribution, we built a simple model to demonstrate a general concept. Hence, we used the simplifying assumption of resource competition theory. Such a simple approach, however, raises new questions that require more complex and system-specific approaches. First, how does spatial and temporal variability affect preemption exploitation? Second, what happens when multiple types of resources are involved in preemption exploitation? Answering these questions is a challenge for future models and empirical tests of the current model.

A central open question is how important is the $R^*$-preemption tradeoff as a coexistence mechanism? In contrast with our initial expectation, we found that partial preemption allows coexistence only under a narrow range. These findings suggest that partial preemption alone is probably not a major mechanism for maintaining biodiversity. Nonetheless, this tradeoff, together with spatial heterogeneity in the abiotic environment, may promote coexistence where different strategies are favored in different microsites.

In contrast with partial preemption, total preemption enables multispecies coexistence in vast parameter space. The assumption of amensalism can be met when species differ in their life form\life history attributes. Under these conditions, we expect the $R^*$-preemption tradeoff to be a crucial coexistence mechanism in various ecosystems. Furthermore, the $R^*$-preemption tradeoff could be a mechanism of character displacement that promotes the divergence of different life forms.

## Data availability statement

No new empirical data was used in this study.

## Competing interests statement

The authors declare no competing financial interests.